\documentclass[12pt,a4paper]{article}
\usepackage{amsfonts}
\usepackage{color,caption}
\usepackage{graphicx}
\usepackage{amsmath}
\usepackage{float}
\usepackage{amssymb}

\begin{document}

\begin{center}
{\Large \bf Central charge for the Schwarzschild black hole} \\
\vskip .5cm
K. Ropotenko\\
\centerline{\it Taras Shevchenko National University of Kyiv}
\centerline{\it 4g Academician Hlushkov Ave.,  Kyiv, 03022, Ukraine}
\bigskip
\verb"ropotenko@ukr.net"

\end{center}
\bigskip\bigskip
\begin{abstract}
Proceeding in exactly the same way as in the derivation of the
temperature of a dual CFT for the extremal black hole in the
Kerr/CFT correspondence, it is found that the temperature of a
chiral, dual CFT for the Schwarzschild black hole is $T=1/2\pi$.
Comparing Cardy's formula with the Bekenstein-Hawking entropy and
using $T$, it is found that the central charge for the Schwarzschild
black hole is of the form $c=12J_{\rm in}$, where $J_{\rm in}$ is
the intrinsic angular momentum of the black hole, $J_{\rm in}=A/8\pi
G$. It is shown that the central charge for any four-dimensional
(4D) extremal black hole is of the same form. The possible
universality of this form is briefly discussed.

\end{abstract}
\bigskip\bigskip

\section{Introduction}

The microscopic origin of the Bekenstein-Hawking relation between
entropy and area of a black hole
\begin{equation}
\label{in1} S_{\rm BH}=\frac{A}{4 l_{\rm P}^{2}}
\end{equation}
remains a central problem in black hole physics. The relation is
universal; it holds for any type of black holes in any dimension.
The universality can be a key to the problem; it implies the
existence of a simple underlying structure. It is now believed that
universal conformal structure of a two-dimensional conformal field
theory (2D CFT) can provide a solution to the problem of the
Bekenstein-Hawking entropy. According to this point of view, the
Bekenstein-Hawking entropy is a representation of the Cardy formula,
a universal relation in statistical mechanics, which relates the
entropy of the CFT to its central charge. In the last years,
considerable progress has been made in reproducing the
Bekenstein-Hawking entropy using the Kerr/CFT correspondence and its
extensions (see reviews \cite{bred} and \cite{comp}). Specifically,
according to the Kerr/CFT correspondence quantum gravity in the
near-horizon extreme Kerr (NHEK) geometry is holographically dual to
a chiral, left-moving $c_L=12J$ 2D CFT at the temperature
$T_L=1/2\pi$, where $J$ is the angular momentum of an extremal Kerr
black hole. Then the Bekenstein-Hawking entropy $S_{\rm BH}=2\pi J$
can be reproduced by the Cardy formula
\begin{equation}
\label{in2} S_{\rm C} = \frac{\pi^{2}}{3}c_LT_L.
\end{equation}
The Kerr/CFT correspondence can be further generalized to the
extremal Reissner-Nordstr\"{o}m case regarding the $U(1)$ electric
charge as an angular momentum in a higher-dimensional spacetime.
Unfortunately, it is not so easy to extend the Kerr/CFT
correspondence to the non-extremal black holes. The problem is that
away from the extremal limit the NHEK geometry disappears and the
near-horizon geometry is just Rindler space. To date, however, there
are no generally accepted ways to associate a conformal field theory
to the Rindler space.

This problem was circumvented by using the so-called hidden
conformal symmetry. This symmetry is not derived from the conformal
symmetry of spacetime geometry itself, but is probed by the
perturbation fields in the near horizon region. Using the hidden
symmetry and the central charges $c_L$ ($c_R=c_L$) derived at
extremality, it was shown that the Bekenstein-Hawking entropy of the
Kerr, Reissner-Nordstr\"{o}m (RN) and Kerr-Newman (KN) black holes
can be reproduced via the Cardy formula $S_{\rm C} =
\frac{\pi^{2}}{3}(c_LT_L+c_RT_R)$ in a 2D CFT at the temperatures
$T_L$, $T_R$.

Recently, Bertini, Cacciatori and Klemm \cite{bert} showed that the
Schwarzschild black hole also enjoys a hidden conformal symmetry and
might be described by a chiral CFT. Going further in this direction,
Lowe and Skanata \cite{low} assumed that the temperature of the dual
CFT is $T = T_{\rm H}$ and, using the Cardy formula, found that the
central charge for the Schwarzschild black hole is of the form
\begin{equation}
\label{in3} c = 12r_+^{3}.
\end{equation}

Despite the considerable progress many problems remain \cite{bred},
\cite{comp}. The main problem is that to date there are no
calculations of $c$ for the non-extremal black holes away from
extremality. Therefore, it is unclear how the result (\ref{in3})
agrees with others, since the derivations of $c$ for the Kerr, RN
and KN black holes are done at extremality, which cannot be done in
the Schwarzschild case. The importance of the central charge
consists, in particular, in the fact that it is proportional to the
number of degrees of freedom. If a CFT is holographically dual to a
black hole, this number should be proportional to the black hole
area. In the extremal case all the $c$ satisfy this requirement.
However, in the non-extremal case this is not so. Therefore, it is
unclear how to interpret the central charge in the non-extremal case
and, in particular, the dependence of $c$ on $r_+^{3}$ in
(\ref{in3}) for the Schwarzschild black hole. Moreover, it is
unclear how the temperature $T = T_{\rm H}$ agrees with the regime
of applicability of the Cardy formula, which requires the
temperature $T$ to be large compared to $c$.

In this note we propose an alternative calculation of $c$ for the
Schwarzschild black hole. We do not use the concept of hidden
conformal symmetry. Proceeding in exactly the same way as in the
derivation of $T_L$ for the extremal Kerr black hole in the Kerr/CFT
correspondence, and using the concept of the intrinsic angular
momentum of a black hole $J_{\rm in}$ introduced in \cite{ro1},
\begin{equation}
\label{in4} J_{\rm in}=\frac{A}{8\pi G},
\end{equation}
we get
\begin{equation}
\label{in5} T_L = \frac{1}{2\pi}.
\end{equation}
Next, comparing Cardy's formula with the Bekenstein-Hawking entropy
$S_{\rm BH}=2\pi J_{\rm in}$ and using $T_L$ we get
\begin{equation}
\label{in6} c_L = 12J_{\rm in}=\frac{3A}{2\pi G}.
\end{equation}
This value is the same as obtained by Carlip \cite{car1}. Long
before the Kerr/CFT correspondence Carlip in his ``the horizon as a
boundary" approach found that in any spacetime of dimension greater
than two, the subgroup of diffeomorphisms in the $(r, t)$ plane
becomes a Virasoro algebra with the central charge of the form
(\ref{in6}). Solodukhin \cite{sol}, using a similar near-horizon
approach, also found that the central charge is proportional to the
Bekenstein-Hawking entropy.

Universality of the Bekenstein-Hawking entropy implies a similar
universality of $c_L$. In turn, universality of $c_L$ follows from
that of $J_{\rm in}$. This is a key observation of the paper. Using
this universality as a guide, we show explicitly that parameters of
the extremal 4D  black holes, being expressed in terms of $J_{\rm
in}$, take the same forms as in the Schwarzschild case. We
conjecture that the problem of universality can be resolved if there
is a universal chiral CFT with $c_L$ of the form (\ref{in6}) for any
4D black hole.

This paper is organized as follows. In section 2 we summarize
without proofs the relevant material on $J_{\rm in}$. In section 3
it is shown that the central charge for the Schwarzschild black hole
is $c_L=12J_{\rm in}$, our main result. Section 4 shows that this
formula is valid not only for the Schwarzschild black hole, but also
for the 4D extremal black holes. In section 5 we discuss our
conclusions and conjectures.

\section{Intrinsic angular momentum}

In this section we shall show that any 4D black hole is
characterized not only by its mass $M$, angular momentum $J$, and
electric charge $Q$ but also by its \emph{intrinsic} angular
momentum $J_{\rm in}$.

The concept of $J_{\rm in}$ was first introduced in \cite{ro1}.
Later, it was generalized in the works \cite{med}-\cite{ro4}. For
the convenience of the reader we repeat and summarize the relevant
material from the works without proofs, thus making our exposition
self-contained.

\subsection{Schwarzschild black hole}

The Bekenstein-Hawking entropy is a concept defined in the rest
frame of an external fiducial (fixed $r$) observer. From the point
of view of the observer, the event horizon is the end of space.
Moreover, ordinary time stands still here and effectively
disappears. So does motion. Therefore, we shall regard the Euclidean
formulation as more fundamental; it is an analytic continuation of
that part of the Lorentzian geometry that just lies outside or at
the event horizon $r\geq 2GM$. Further, since the $(r, t)$ part of
the Schwarzschild metric is well approximated, around the horizon,
by flat space, where all our quantities are well-defined, we shall
deal only with the Rindler section of the whole Euclidean
Schwarzschild manifold. I begin with the Schwarzschild metric
\begin{equation}
\label{s1}ds^{2}=-\left(1-\frac{2GM}{r}\right)dt^{2}+
\left(1-\frac{2GM}{r}\right)^{-1}dr^{2}+r^{2}d\Omega^{2}.
\end{equation}
Analytically continuing to imaginary time $t=-it_{\rm E}$, where
$t_{\rm E}$ is a real parameter and the subscript ``E" means
Euclidean, we obtain the Euclidean Schwarzschild metric
\begin{equation}
\label{s2}ds_{\rm E}^{2}=\left(1-\frac{2GM}{r}\right)dt_{\rm E}^{2}+
\left(1-\frac{2GM}{r}\right)^{-1}dr^{2}+r^{2}d\Omega^{2}.
\end{equation}
In the near-horizon approximation it takes the Rindler form
\begin{equation}
\label{s3} ds_{\rm E}\approx \rho^{2}d(kt_{\rm
E})^{2}+d\rho^{2}+\frac{1}{4k^{2}}d\Omega^{2},
\end{equation}
where $\rho$ is the proper distance from the horizon,
\begin{equation}
\label{s4} \rho=\int_{r_+}^{r} \sqrt{g_{rr}(r')}dr',
\end{equation}
and $k$ is the surface gravity of a Schwarzschild black hole, $k =
1/4GM$. It is the product of the metric on a two-sphere of radius
$2GM$ (the last term) and the metric of $(\rho, t_{\rm E})$ plane
\begin{equation}
\label{s5}ds_{\rm E}=\rho^{2}d(kt_{\rm E})^{2}+d\rho^{2}.
\end{equation}
Evidently we have invariance under rotation generated by the Killing
operator $\partial/\partial t_{\rm E}$ in the plane and the
two-sphere $r_+=2GM$ is an axis of rotation. The $(\rho, t_{\rm E})$
plane is smooth there if $\Theta_{\rm E}=kt_{\rm E}$ is an angular
coordinate with period $2\pi$; $t_{\rm E}$ itself has then
periodicity $\beta=2\pi/k$, the inverse of the Hawking temperature.
This is the geometrical origin of the universal character of the
Hawking temperature.

If there is rotation associated with $\Theta_{\rm E}$, then,
according to quantum mechanics, there is an angular momentum
component (say, the $z^{\rm{th}}$, here denoted by $\hat{J}_{\rm
in}$) conjugate to $\Theta_{\rm E}$, $\hat{J}_{\rm in}=-i\hbar
\partial/\partial\Theta_{\rm E}$. $\hat{J}_{\rm in}$ is, apart from a
factor, the same as the Killing operator $\partial/\partial t_{\rm
E}$. Below, we shall often speak of the ``angular momentum",
understanding by this the eigenvalue $J_{\rm in}$,
\begin{equation}
\label{s6} -i\hbar \frac{\partial \psi}{\partial \Theta_{\rm
E}}=J_{\rm in}\psi.
\end{equation}
There are a number of ways to calculate $J_{\rm in}$. One heuristic
way is to derive $J_{\rm in}$ from the action integral for a black
hole. According to the Euclidean path integral approach to quantum
gravity all the thermodynamical properties of a black hole are
completely determined by the partition function $Z$, which in turn
is determined, in the near-horizon approximation, by the Euclidean
action integral for a black hole $\ln Z\approx iI=-I_{\rm E}$.
Gibbons and Hawking \cite{gib} found that for the Schwarzschild
black hole
\begin{equation}
\label{s7} I_{\rm E}=\frac{1}{2}\beta M.
\end{equation}
On the other hand, since motion of the black hole in Euclidean
sector has a character of rotation, the action integral is
determined by $J_{\rm in}$,
\begin{equation}
\label{s8} I_{\rm E}=\oint J_{\rm in}\, d\Theta_{\rm E} =2\pi J_{\rm
in}.
\end{equation}
Equating these two expressions, we get
\begin{equation}
\label{s9} J_{\rm in} =2GM^{2}=\frac{A}{8\pi G}.
\end{equation}
We can obtain the same result more rigorously using the commutator
\cite{ro1}
\begin{equation}
\label{s10} [J_{\rm in}, \hat{\Theta}_{\rm E}]=-i\hbar,
\end{equation}
where the operator $\hat{\Theta}_{\rm E}$ is defined as
$\hat{\Theta}_{\rm E}=k\hat{t}_{\rm E}$. However, as Medved
\cite{med} noted, this approach is rather complicated because it
involves the Euclidean operator $\hat{\Theta}_{\rm E}$. Conversely,
Medved pointed out that $\Theta_{\rm E}$ and $A/8\pi G$, just as
$t_{\rm E}$ and $M$, appear as canonical conjugates in the off-shell
black hole action of Bunster (Teitelboim) and Carlip \cite{car2}
\begin{equation}
\label{s11} I_{\rm E} =-\Theta_{\rm E}\left(\frac{A}{8\pi
G}\right)+I_{can}-t_{\rm E}M.
\end{equation}
Here all the variables are well-defined, so that the result
(\ref{s9}) follows from the action immediately without any technical
problems. Note that similarly to the derivation of Schr\"{o}dinger's
equation from the variational principle, Bunster and Carlip derived
from the action (\ref{s11}) the Schr\"{o}dinger-type equation
\cite{car2}
\begin{equation}
\label{s12} \frac{\hbar}{i}\frac{\partial \psi}{\partial
\Theta}-\frac{A}{8\pi G}\psi=0.
\end{equation}
It is evident that this equation gives the same result. Indeed, in
the semiclassical approximation
\begin{equation}
\label{s13} \psi=a\exp \left(-\frac{i}{\hbar}I\right),
\end{equation}
where $I$ is the action of a black hole. Substituting this in
(\ref{s12}) we obtain
\begin{equation}
\label{s14} \frac{\partial I}{\partial \Theta}\psi=\frac{A}{8\pi
G}\psi;
\end{equation}
the slowly varying amplitude $a$ need not be differentiated. Under
Euclidean continuation $\Theta\rightarrow -i\Theta_{\rm E}$ and
$I\rightarrow iI_{\rm E}$,
\begin{equation}
\label{s15}  \frac{\partial I_{\rm E}}{\partial \Theta_{\rm E}
}\psi=\frac{A}{8\pi G}\psi.
\end{equation}
The derivative $\partial I_{\rm E}/\partial \Theta_{\rm E}$ is just
a generalized momentum corresponding to the angle of rotation about
one of the axes (say, the $z^{\rm{th}}$). Therefore $A/8\pi G$ is
what corresponds in quantum mechanics to $J_{\rm in}$. In fact, this
conclusion can be recovered even more simply: analytically
continuing $\Theta$ and $A/8\pi G$ to the imaginary values
$\Theta\rightarrow -i\Theta_{\rm E}$ and $(A/8\pi G)\rightarrow
i(A/8\pi G)_{\rm E}$ with $(A/8\pi G)_{\rm E}=(A/8\pi G)$ in
(\ref{s12}), we immediately obtain the same result.

\subsection{Area quantization}

In classical mechanics, the $z$-component of the angular momentum is
an adiabatic invariant. $J_{\rm in}$ is the $z$-component of the
angular momentum. Since $J_{\rm in}$ is an adiabatic invariant, then
the black hole area is also an adiabatic invariant. So is the
entropy. According to the semiclassical Bohr-Sommerfeld quantization
rule
\begin{equation}
\label{a1} I_{\rm E}=\oint J_{\rm in}\, d\Theta_{\rm E} =2\pi \hbar
\cdot m,\quad m=0,1,2,...\,.
\end{equation}
Therefore,
\begin{equation}
\label{a2} J_{\rm in}=m \hbar.
\end{equation}
In quantum mechanics $m$ can also take negative values related with
rotation around $z$ axis in the negative direction; these are
however associated with the negative surface gravity and can be
rejected. From (\ref{a2}) it follows that the black hole area and
entropy are quantized
\begin{equation}
\label{a3} A=\Delta A \cdot m,
\end{equation}
\begin{equation}
\label{a4} S_{\rm BH}=2\pi m,
\end{equation}
where
\begin{equation}
\label{a5} \Delta A = 8\pi l_P^{2}
\end{equation}
is the quantum of area. This approach was extended to generic
theories of gravity by Medved \cite{med} and to de Sitter space - by
Jia, Mao and Ren \cite{mao}. Forty years ago, by proving that the
black hole horizon area is an adiabatic invariant, Bekenstein
\cite{bek1}, \cite{bek2}  showed that the area spectrum of a black
hole is of the form (\ref{a3}). But he did not use the concept of
the intrinsic angular momentum.

Quantization of the black hole area is an important concept because,
as believed, the quantum number $m$ determines the number of degrees
of freedom of a black hole. Moreover, this concept explains
universality of the Bekenstein-Hawking relation, i.e.
proportionality of the entropy and horizon area. Indeed, if the
horizon surface consists of $m$ independent patches of area $8\pi
l_P^{2}$, $m=A/8\pi l_P^{2}$, and every patch-degree of freedom has
$k$ states available to it, then the total number of states is
$k^{m}$ and $S_{\rm BH}\propto A$.

\subsection{Regge trajectories}

The Schwarzschild black hole is unstable and decays by emitting
Hawking radiation. Therefore, it can be viewed as a resonance and
$J_{\rm in}$ as its angular momentum. It is interesting that $J_{\rm
in}$ is proportional to the square of the mass of a black hole
\begin{equation}
\label{re1} J_{\rm in}=2G M^{2}.
\end{equation}
This resembles the well-known angular momentum-mass relation for
hadronic resonances. As is well known, the graph of the angular
momentum $J$ of hadronic resonances against their mass squared falls
into lines $J=\alpha' M^{2}$ called Regge trajectories. Instead of
terminating abruptly as in the case of nuclei, the graph continue on
indefinitely, implying that quarks don't fly apart when spun too
fast. In contrast to nuclei, there is no a threshold there. This is
a manifestation of quark confinement. Analogously, we can assume
that the black holes lie on the Regge trajectories with the slope
$\alpha'=2G$. Since $J_{\rm in}$ increases with the square of the
black hole mass without limit, the black hole area never decreases,
just as the area theorem predicts. This is a manifestation of
gravitational confinement.

\subsection{Kerr-Newman black hole}

The intrinsic angular momentum for the Kerr, RN and KN black holes
can be defined in exactly the same way as in the Schwarzschild case
\cite{ro1}. Without repeating the calculations, we give the result:
\begin{equation}
J_{\rm in} \equiv \frac{A}{8 \pi G}=\left\{
\begin{array}{lll}
2GM^{2},& \mbox{Sch}\\
    GM^{2}+M\sqrt{G^{2}M^{2}-a^{2}}, & \mbox{Kerr}\\
    GM^{2}-Q^{2}/2 +M\sqrt{G^{2}M^{2}-GQ^{2}}, & \mbox{RN}\\
    GM^{2}-Q^{2}/2+M\sqrt{G^{2}M^{2}-GQ^{2}-a^{2}}, &
    \mbox{KN}.
  \end{array}
  \right.
\label{kn1}
\end{equation}
where $Q$ is the electric charge of a black hole and $a$ is the
specific angular momentum, $a=J/M$ ($J$ being the angular momentum).

With $J_{\rm in}$, the entropy of any black hole takes the form
\begin{equation}
\label{kn2} S_{\rm BH}=2\pi J_{\rm in},
\end{equation}
and all terms in the first law of black hole mechanics look more
uniformly,
\begin{equation}
\label{kn3} dM=kdJ_{\rm in}+\Omega_H dJ+\Phi_H dQ.
\end{equation}
It is believed that the laws of black hole mechanics have no
independent physical significance and acquire it only after
identifying with the laws of thermodynamics. However, if the concept
of the intrinsic angular momentum is correct, the first term in
(\ref{kn3}) can have a direct physical interpretation: it is the
change in the black hole energy due to rotation in internal space.
Therefore, the first law of black-hole mechanics can have a
mechanical meaning.

\section{Temperature, conformal weight, and central charge}

In this section, we shall calculate the central charge for the
Schwarzschild black hole. Following Bertini, Cacciatori and Klemm
\cite{bert} and also Lowe and Skanata \cite{low} we assume that the
Schwarzschild black hole is nothing but a thermal state of a chiral
$c_L$ $(c_R=0)$ CFT. In the canonical ensemble, the entropy of the
thermal state is given by the Cardy formula
\begin{equation}
\label{b1} S_{\rm C} = \frac{\pi^{2}}{3}c_L\,T_L,
\end{equation}
where $T_L$ is the temperature of the state. The appearance of
temperature means that there is periodic evolution and the
background topology is a cylinder of circle $1/T_L$, not a complex
plane. Then the entropy in the microcanonical ensemble is given by
the Cardy formula
\begin{equation}
\label{b2} S_{\rm C} = 2\pi
\sqrt{\frac{c_L}{6}\left(L_0-\frac{c_L}{24}\right)},
\end{equation}
where $L_0$ is the conformal weight, the eigenvalue of the zero-mode
Virasoro operator, related to the $T_L$ by the first law of
thermodynamics.

We shall calculate $c_L$ by comparing Cardy's formula (\ref{b1}) (or
(\ref{b2})) with the Bekenstein-Hawking entropy. For this purpose we
need $T_L$ (or $L_0$).

\subsection{Temperature and
conformal weight}

It is believed that the Schwarzschild case is more difficult than
others because the derivations of the central charges of a dual CFT
for the Kerr, RN or KN black holes are done at extremality, which
cannot be done in the Schwarzschild case \cite{comp}. However,
although the Schwarzschild black hole has no extremal limit, there
is an analogy between the Schwarzschild and the 4D extremal black
holes. This analogy is based on the fact that for the extremal just
as for the Schwarzschild black hole the mass $M$ is the only
parameter that completely specifies the metric. As will be
demonstrated later, the analogy becomes even more striking if we use
the concept of the intrinsic angular momentum $J_{\rm in}$. In the
case of an extremal black hole the temperature of a CFT, i.e. the
thermodynamic potential dual to the zero mode of the Virasoro
algebra, follows from the first law of black hole thermodynamics in
the near horizon region \cite{har}. To calculate $T_L$ (or $L_0$)
for the Schwarzschild black hole we proceed in the same way as in
the case of an extremal black hole. The Hartle-Hawking vacuum state
around a Schwarzschild black hole is characterized by the Boltzmann
factor at the Hawking temperature $T_{\rm H}$. However, in the near
horizon region we have the Rindler space and the Rindler vacuum.
Like the near horizon region of an extremal black hole, the Rindler
space of the Schwarzschild black hole can be treated as an isolated
geometry with its own thermodynamics. By analogy with thermodynamics
of the near horizon region of an extremal black hole \cite{har}, the
first law of thermodynamics of the Rindler space can be written in
the form
\begin{equation}
\label{t1} dS_{\rm BH}=\frac{dJ_{\rm in}}{T_L},
\end{equation}
where $J_{\rm in}$ and $T_L$ are the energy and temperature of the
Rindler space. From this we get
\begin{equation}
\label{t2} T_L=\frac{1}{2\pi}.
\end{equation}
By analogy with the Kerr/CFT correspondence, we assume that the near
horizon region of the Schwarzschild black hole, the Rindler space,
is holographically dual to a chiral CFT at the left-moving
temperature (\ref{t2}) which is conjugate to the zero mode of the
Virasoro algebra. Similarly, the vacuum state of quantum fields in
the Rindler space, the Rindler vacuum, is characterized by the
Boltzmann factor of the form
\begin{equation}
\label{t3} \exp\left(-\frac{L_0}{T_L}\right ),
\end{equation}
where $L_0$ is the eigenvalue of the Virasoro operator, the
conformal weight,
\begin{equation}
\label{t4} L_0=J_{\rm in}.
\end{equation}

\subsection{Central charge}

Comparing now the Cardy formula (\ref{b1}) with the
Bekenstein-Hawking entropy $S_{\rm BH}=2\pi J_{\rm in}$ and using
the temperature (\ref{t2}), we find
\begin{equation}
\label{ch1} c_L=12J_{\rm in}=\frac{3A}{2\pi G}.
\end{equation}
Comparing the Cardy formula (\ref{b2}) with the Bekenstein-Hawking
entropy $S_{\rm BH}=2\pi J_{\rm in}$ and using the the conformal
weight (\ref{t4}) we get the same result. By definition, the zero
mode Virasoro operators are linear combinations of dilations and
rotations, so that the corresponding eigenvalues are linear
combinations of the ordinary ADM mass and intrinsic angular
momentum:
\begin{equation}
\label{ch3} L_0=\frac{Mr_++J_{\rm in}}{2},\qquad
\bar{L}_0=\frac{Mr_+-J_{\rm in}}{2}.
\end{equation}
Since we have a $c_L$ $(c_R=0)$ chiral CFT, the shifted values are
\begin{equation}
\label{ch4} L_0-\frac{c_L}{24}=\frac{J_{\rm in}}{2},\qquad
\bar{L}_0-\frac{c_R}{24}=0.
\end{equation}

\section{Central charges for the extremal black holes}

It turns out that with $J_{\rm in}$ the parameters of a dual CFT for
the extremal black hole take the same forms as those for the
Schwarzschild black hole. Let us consider this fact in more detail.
According to the Kerr/CFT correspondence and its extensions a 4D
extremal black hole is holographically dual to a chiral, left-moving
half of a $c_L=c_R$ 2D CFT at the temperature $T_L$ ($T_R=0$) (in
the case of an extremal KN black hole there are two different CFT
holographically dual to the KN black hole, so that $c_L$ is a linear
combination of two limiting cases corresponding to the Kerr/CFT and
RN/CFT correspondences). The parameters of the CFTs, taken from
\cite{comp}, are given in Table 1. Here, $R_\chi$ is the radius of
the Kaluza-Klein circle in fifth dimension.
\begin{center}
\captionof{table}{Parameters of the dual CFTs (from \cite{comp})}
\label{tab:tab1}
\begin{tabular}{ | l | l | l| l | l |}
\hline & extremal Kerr & extremal RN
\\ \hline
central charge, $c_L$ & $12J$ &  $6Q^{3}\sqrt{G}/R_\chi$ \\
\hline temperature, $T_L$ & $1/2\pi$ & $R_\chi/2\pi\sqrt{G} Q$ \\
\hline entropy, $S_{\rm C}$ & $2\pi J$ & $\pi Q^{2}$
\\\hline
\end{tabular}
\end{center}
\bigskip
In general, $R_\chi$ is a free parameter. However, Chen \emph{et al}
\cite{chen1}, \cite{chen2} noted that
\begin{equation}
\label{e1} R_\chi=Q\sqrt{G}
\end{equation}
is the most reasonable choice (in particular, the near horizon
geometry of the five-dimensional (5D) uplifted RN black hole is then
exactly $AdS_3\times S^{2}$).  Using this choice, let us rewrite the
parameters of CFTs in terms of $J_{\rm in}$. Since the surface
gravity $k=0$, and hence $\Theta_{\rm E}=0$, one cannot determine
the extremal $J_{\rm in}$ as was done in the non-extremal case.
However, it is still possible to determine the extremal $J_{\rm in}$
as a value of $J_{\rm in}$ at extremality. For an extremal Kerr
black hole we get
\begin{equation}
\label{e2} J_{\rm in}\equiv\frac{A_{\rm ext}}{8 \pi G}=GM^{2}
\end{equation}
and the intrinsic angular momentum becomes an ordinary spin
\begin{equation}
\label{e3} J_{\rm in}=J.
\end{equation}
Then
\begin{equation}
\label{e4} c_L=12J_{\rm in}.
\end{equation}
For an extremal RN black hole
\begin{equation}
\label{e5} J_{\rm in}\equiv\frac{A_{\rm ext}}{8\pi
G}=\frac{Q^{2}}{2}
\end{equation}
and the central charge takes the same form (\ref{e4}). The
temperature of a dual CFT for the extremal black hole follows from
the first law of black hole thermodynamics,
\begin{equation}
\label{e6} dS_{\rm BH}=\frac{dJ_{\rm in}}{T_L},
\end{equation}
and is the same in both cases
\begin{equation}
\label{e7} T_L=\frac{1}{2\pi}.
\end{equation}
The parameters of the Schwarzschild and extremal black holes are
summarized in Table 2 for comparison.
\begin{center}
\captionof{table}{Summary of the CFT parameters} \label{tab:tab1}
\begin{tabular}{ | l |c | c | c|}
\hline & Schwarzschild & extremal Kerr & extremal RN
\\ \hline
intrinsic angular momentum, $J_{\rm in}$ & $2G M^{2}$  &  $J$ &  $Q^{2}/2$ \\
\hline
central charge, $c_L$ & $12J_{\rm in}$ &  $12J_{\rm in}$ &  $12J_{\rm in}$ \\
\hline temperature, $T_L$ & $1/2\pi$ & $1/2\pi$ & $1/2\pi$ \\
\hline entropy, $S_{\rm C}$ & $2\pi J_{\rm in}$  & $2\pi J_{\rm in}$
& $2\pi J_{\rm in}$
\\\hline
\end{tabular}
\end{center}
\bigskip

As mentioned above, the extremal $J_{\rm in}$ cannot be defined in
the same way as the non-extremal one. Therefore, unlike the
non-extremal case, the extremal $J_{\rm in}$ is not quantized as an
integer multiple of $\hbar$. Indeed, although in the case of an
extremal Kerr black hole
\begin{equation}
\label{e8} J_{\rm in}=\frac{A_{\rm ext}}{8 \pi G}=J = m\hbar ,
\end{equation}
where $m$ is integer or half integer, in the case of an extremal RN
black hole this is not so:
\begin{equation}
\label{e9} J_{\rm in}=\frac{A_{\rm ext}}{8 \pi G}=\frac{Q^{2}}{2}
\neq m\hbar.
\end{equation}
The last result is evident: since the fine structure constant
$\alpha$ is not an even integer, $\alpha=e^{2}/\hbar c\simeq 1/137$,
the quantity $Q^{2}/2$, where the electric charge $Q$ is quantized
in terms of the electron charge $e$, is not an integer multiple of
$\hbar$. According to the theory of the renormalization group, the
value of $\alpha$ grows logarithmically as the energy scale is
increased; the usual value $\alpha=e^{2}/\hbar c\simeq 1/137$ is
defined as the square of the completely screened charge, that is,
the value observed at infinite distance or in the limit of zero
momentum transfer. What happens if we get closer to the black hole?
According to the RN/CFT correspondence the near horizon region of an
extremal RN black hole is holographically described by a CFT with
parameters depending on the radius of the Kaluza-Klein circle in
fifth dimension $R_\chi$. According to the Kaluza-Klein theory
\cite{zee}, the electric charge $Q$ in four dimension is
proportional to the momentum of the motion round the curled up fifth
dimension $P$,
\begin{equation}
\label{e10} Q=P \sqrt{G}.
\end{equation}
Since $P$ is quantized
\begin{equation}
\label{e11} P=\frac{m\hbar}{R_\chi},\quad m\in\mathbb{Z},
\end{equation}
$Q$ is also quantized
\begin{equation}
\label{e12} Q=\frac{ m\hbar\sqrt{G}}{R_\chi}.
\end{equation}
Using the choice (\ref{e1}), we get
\begin{equation}
\label{e13} Q^{2}= m\hbar.
\end{equation}
Therefore, (\ref{e9}) becomes an equality for half-integer $m$.

In the original version of the Kaluza-Klein theory the masses of
charged particles are simply the momenta along the extra dimension
and are quantized in units of $1/R_\chi$. There is however a serious
difficulty with this prescription: the masses come out to be of the
order of the Planck mass. This means that these particles do not
correspond to the observed charged particles. We now turn attention
to the fact that the mass of any black hole is always greater than
or equal to the Planck mass. Moreover, the mass of an extremal RN
black hole is of the form
\begin{equation}
\label{e14} M_m=\frac{Q^{2}}{r_+}=\frac{2J_{\rm
in}}{r_+}=\frac{|m|\hbar}{r_+},
\end{equation}
where, according to (\ref{e1}), the radius of the black hole is
equal to that of the extra dimension
\begin{equation}
\label{e15} r_+=R_\chi.
\end{equation}
Note also that like a Kaluza-Klein particle, an extremal RN black
hole satisfies the Bogomol'nyi identity
\begin{equation}
\label{e16} M=|Q|G^{-1/2}.
\end{equation}
Therefore, we expect that the above difficulty can be overcome in
part if we identify the Kaluza-Klein particles with the excitations
of the near horizon region of the extremal RN black holes. It is
interesting that the mass of any 4D black hole (whether or not
extremal) can also be written in the Kaluza-Klein form
\begin{equation}
\label{e17} M_m=\frac{J_{\rm in}}{r_+}=\frac{|m|\hbar}{r_+},\quad
\mbox{(Schwarzschild, Kerr)},
\end{equation}
or
\begin{equation}
\label{e18} M_m=\frac{J_{\rm
in}+\frac{Q^{2}}{2}}{r_+}=\frac{|m|\hbar}{r_+},\quad\mbox{(RN, KN)}.
\end{equation}

\section{Discussion}

In this paper we have calculated the central charge $c_L$ for the
Schwarzschild black hole. We have found that $c_L=12J_{\rm in}$,
where $J_{\rm in}$ is the intrinsic angular momentum of the black
hole, $J_{\rm in}=A/8\pi G$. We have also found the temperature of a
dual CFT $T_L=1/2\pi$ and the conformal weights $L_0=(Mr_++J_{\rm
in})/2=J_{\rm in}$ and $\bar{L}_0=(Mr_+-J_{\rm in})/2=0$.

Our result means that $c$ is proportional to the black hole area,
not volume. This conclusion is important because it agrees with the
meaning of $c$. Generally $c$ is proportional to the number of
degrees of freedom of a CFT. According to the Kerr/CFT
correspondence and its extensions the degrees of freedom of a 4D
black hole reside on its horizon. If the CFT is holographically dual
to a black hole, that number should be proportional to the black
hole area. Therefore, $c$ should be proportional to the black hole
area, not volume. This disagrees with the result of Lowe and
Skanata; they found that $c$ is proportional to the black hole
volume (\ref{in3}). The reason for this is that they used $T =
T_{\rm H}$ and an indirect method to calculate $c$. We also
calculated $c$ in an indirect way. However, our result agrees with
that of Carlip \cite{car1} and Solodukhin \cite{sol} obtained by
direct calculation.

The advantage of our approach is that it does not need the concept
of hidden conformal symmetry; we have calculated $T_L $ proceeding
in exactly the same way as in the derivation of the temperature of a
dual CFT for the extremal black hole in the Kerr/CFT correspondence.
We have found that $T_L $ is nothing but the Rindler temperature
$T_{Rindler}=1/2\pi$. This result is important because it justifies
the high-temperature regime of applicability of the Cardy formula.
Indeed, a local proper temperature measured by a Rindler observer at
distance $\rho$ from the horizon can be obtained from the Rindler
temperature by using the transformation between Rindler and proper
time. The local temperature is thus given by $T_{local}=1/2\pi\rho
=T_{Rindler}/\rho$ and increases as we move toward the horizon.

Although our approach does not require a detailed knowledge of a CFT
and its Virasoro algebra, we expect that, contrary to previous
views, this Virasoro arises not from an enhancement of the $SL(2;R)$
symmetry appearing in the scalar wave equation in the Schwarzschild
background but rather from the $U(1)$ rotation isometry generated by
$J_{\rm in}$ in the near horizon Rindler space. Since we have only
one $U(1)$, we have a chiral CFT.

The importance of our result is due to the fact that, as we have
demonstrated above, all the CFT parameters of the extremal black
holes, including $c_L$, being written in terms of $J_{\rm in}$, take
the same forms as those for the Schwarzschild black hole. The reason
is that the Schwarzschild just as the extremal black hole contains
$U(1)$ rotation isometry generated by $J_{\rm in}$ in the
near-horizon geometry and the mass $M$ is the only parameter that
completely specifies its metric. Therefore, the form of $c_L$ is, at
least in part, universal. This can help in understanding
universality of the Bekenstein-Hawking relation.

Can one extend this universality to the non-extremal Kerr, RN and KN
black holes? The existence of holographically dual CFTs to
non-extremal black holes is highly conjectural \cite{comp}. To date
there are no calculations of $c_L$ for these black holes away from
extremality; their Bekenstein-Hawking entropies are reproduced via
the Cardy formula in CFTs with central charges derived at
extremality. On the other hand, the metric of any non-extremal black
hole can be reduced to the Rindler form with the same $U(1)$
isometry generated by $J_{\rm in}$. Since the entropy of any black
hole (whether or not extremal) depends only on the intrinsic angular
momentum $S_{\rm BH}=2\pi J_{\rm in}$, we can define the chemical
potential $1/T_L=dS_{\rm BH}/dJ_{\rm in}$. Using universality as a
guide, it is natural to conjecture that any 4D black hole is dual to
a chiral $c_L=12J_{\rm in}$ CFT at the temperature $T_L=1/2\pi $. We
hope that future investigation will help to support or falsify our
conjecture.

\section*{Note Added in Proof}

The central charges for the Schwarzschild and other 4D black holes
can be obtained in more simple way if we assume that there exists a
close relation between the central charge and the parameters of the
$AdS_2$ component of the near horizon geometry. The near horizon
geometry of all extreme black holes has an $AdS_2$ factor. Chen
\emph{et al} \cite{chen3} noted that the central charges for the
extremal Kerr and RN black holes satisfy the relation
$c=6\ell^{2}/G$, where $\ell$ is the $AdS_2$ radius of the near
horizon geometry. Indeed, in the case of the extremal Kerr black
hole $\ell^{2}=2GJ$ and $c=12J =12J_{\rm in} $. In the case of the
extremal RN black hole $\ell^{2}=GQ^{2}$ and $c=6Q^{2}=12J_{\rm
in}$. The above relation is valid also for the extremal KN black
hole. In this case $\ell^{2}=r_+^{2}+a^{2}$ and $c=12J_{\rm in}$,
where $J_{\rm in}=A/8\pi G$. It turns out that the near horizon
geometry of the Schwarzschild black hole also has an $AdS_2$
component. Bertini \emph{et al} \cite{bert} showed that the
Schwarzschild solution can be mapped into the near horizon limit
$AdS_2\times S^{2}$ of the extremal RN black hole by a Kinnersley
transformation. In this case $\ell^{2}=4G^{2}M^{2}$. If the above
relation is assumed valid for the Schwarzschild black hole, we have
immediately $c=3A/2\pi G =12J_{\rm in}$.

\end{document}